\documentclass[twocolumn,
aps,showpacs,preprintnumbers,amsmath,amssymb,floatfix,prl]{revtex4}

\usepackage{color,amsmath,graphicx,pst-text,%
amssymb,
            pst-grad,pstricks,pst-3d%,optima
}
\usepackage{leftidx}
\input epsf
\usepackage{epsfig}
\begin{document}

\newcommand{\alphaFn}[1]{\alpha\left(#1\right)}
\newcommand{\Laplace}[2]{\mathcal{L}\left\lbrace #1\right\rbrace\left(#2\right)}
\newcommand{\Mellin}[2]{\mathcal{M}\left\lbrace #1\right\rbrace\left(#2\right)}
\newcommand{\Fourier}[2]{\mathcal{F}\left[ #1\right]\left(#2\right)}
\newcommand{\FourierC}[2]{\mathcal{F}_{c}\left[ #1\right]\left(#2\right)}
\newcommand{\Fourierb}[2]{\mathcal{F}\left\lbrace #1\right\rbrace\left(#2\right)}
\newcommand{\FourierCb}[2]{\mathcal{F}_{c}\left\lbrace #1\right\rbrace\left(#2\right)}
\newcommand{\Fracderiv}[1]{\mathcal{D}^{1-#1}}
\newcommand{\FracIntN}[1]{\frac{\partial^{#1}}{\partial t^{#1}}}
\newcommand{\FracIntM}[2]{\frac{\partial^{#1} #2 }{\partial t^{#1}}}
\newcommand{\FracderivN}[1]{\frac{\partial^{1-#1}}{\partial t^{1-#1}}}
\newcommand{\FracderivM}[2]{\frac{\partial^{1-#1} #2 }{\partial t^{1-#1}}}
\newcommand{\RLderivO}[1]{\frac{d^{#1}}{dt^{#1}}}
\newcommand{\RLderivOM}[2]{\frac{d^{#1} #2}{dt^{#1}}}
\newcommand{\del}{\partial}

\title{Fractional Fokker-Planck Equations for Subdiffusion with Space-and-Time-Dependent Forces.}

\author{B.I. Henry}
\email{B.Henry@unsw.edu.au}
\affiliation{Department of Applied Mathematics, School of Mathematics and Statistics,
University of New South Wales, Sydney NSW 2052, Australia.}

\author{T.A.M. Langlands}
\email{t.langlands@usq.edu.au}
\affiliation{Department of Mathematics and Computing,
University of  Southern Queensland, Toowoomba Queensland 4350, Australia.}

\author{P. Straka}
\affiliation{School of Mathematics and Statistics,
University of New South Wales, Sydney NSW 2052, Australia.}

\date{\today}
\begin{abstract}
We have derived a fractional Fokker-Planck equation for subdiffusion in a general space-and-time-dependent force field from power law waiting time continuous time random walks biased by Boltzmann weights. The governing equation is derived from a generalized master equation  and is shown to be equivalent to a subordinated stochastic Langevin equation.  
\end{abstract}
\keywords{Anomalous subdiffusion, Fractional Fokker-Planck}
\pacs{05.40.Fb,02.50.Ey,05.10.Gg}
\maketitle
%\section{Introduction}

Over the past few decades there has been an enormous growth in the numbers of papers devoted to experimental and theoretical aspects of anomalous diffusion \cite{MK2000,HLS2010}. The  landmark review
by Metzler and Klafter in 2000 \cite{MK2000} has been particularly influential,  promoting the description of anomalous diffusion within the framework of continuous time random walks (CTRWs) and fractional calculus.   
There are now numerous applications utilizing this approach in physics, chemistry, biology and finance \cite{HLS2010}.

A central theoretical result in this research was the derivation \cite{MBK1999b,BMK2000} of  a fractional Fokker-Planck (Smoluchowksi) equation \cite{MBK1999a}
\begin{equation}\label{xffp}
\frac{\partial P}{\partial t}=  \leftidx{_0}{D_t^{1-\alpha}}
\left[\kappa_\alpha\frac{\partial^2}{\partial x^2}-\frac{1}{\eta_\alpha}\frac{\partial}{\partial x}
F(x)\right]P(x,t)
\end{equation}
 to describe the evolution of the probability density function $P(x,t)$ for subdiffusion in an external space-dependent force field $F(x)$. In this equation,
 \begin{equation}
\leftidx{_0}{ D_t^{1-\alpha}} Y(t)=\frac{1}{\Gamma(\alpha)}\frac{\partial}{\partial t}
 \int_0^t \frac{Y(t')}{(t-t')^{1-\alpha}}\, dt'
 \end{equation}
 is the Riemann-Liouville fractional derivative, $\kappa_\alpha$ is a fractional diffusion coefficient, $\eta_\alpha=(\beta \kappa_\alpha)^{-1}$
is a fractional friction coefficient, and $\beta$ is the inverse temperature $k_BT$. The fractional Fokker-Planck equation, Eq.(\ref{xffp}), was derived from the continuous time random walk  model of Montroll and Weiss \cite{MW1965}, with power law waiting times \cite{MBK1999b,BMK2000}. 

More recently a modified fractional Fokker-Planck equation,
\begin{equation}\label{tffp}
\frac{\partial P}{\partial t}= 
\left[\kappa_\alpha\frac{\partial^2}{\partial x^2}-\frac{1}{\eta_\alpha}\frac{\partial}{\partial x}
F(t)\right]\,\leftidx{_0}{D_t^{1-\alpha}}P(x,t),
\end{equation}
was derived from power law waiting time CTRWs, using a generalized master equation  \cite{SK2006}, for subdiffusion in a time-dependent force field $F(t)$.
The modified fractional Fokker-Planck equation, Eq.(\ref{tffp}), was also derived for subdiffusion in dichotomously alternating force fields \cite{HPGH2007,HPGH2009}, $F(x)\xi(t)$ with $\xi(t)=\pm 1$, but  a fractional Fokker-Planck equation
to model subdiffusion in general space-and-time-dependent force fields
$F(x,t)$  has remained elusive \cite{HPGH2007,WMW2008,HPGH2009,KJ2010}. 
On the other hand, a subordinated stochastic Langevin equation
has been proposed for modelling
subdiffusion in space-and-time-dependent force fields \cite{WMW2008}.
More recently \cite{MWK2008}, in the case of time-dependent forces, the moments of the stochastic process defined by this stochastic Langevin equation were  shown to coincide with the moments of the modified
fractional Fokker-Planck equation, Eq.(\ref{tffp}).

There have been numerous papers on fractional Fokker-Planck equations in recent years relying on
{\em ad hoc} or phenomenological models \cite{KYC2006,HPGH2007,MK2009,KJ2010}.

In this letter we derive the fractional Fokker-Planck equation,
\begin{equation}\label{xtffp}
\frac{\partial P}{\partial t}= 
\left[\kappa_\alpha\frac{\partial^2}{\partial x^2}-\frac{1}{\eta_\alpha}\frac{\partial}{\partial x}
F(x,t)\right]\,\leftidx{_0}{D_t^{1-\alpha}}P(x,t),
\end{equation}
from power law waiting time CTRWs, using a generalized master equation,  for subdiffusion in a space-and-time-dependent force field $F(x,t)$.
This fractional Fokker-Planck equation is shown to be formally equivalent to
the subordinated stochastic Langevin equation in \cite{WMW2008} for space-and-time-dependent forces.
We also show that the original fractional Fokker-Planck equation,
Eq.(\ref{xffp}), generalized by replacing $F(x)$ with $F(x,t)$, can be recovered from power law waiting time CTRWs in an {\it ad-hoc} generalization of the CTRW
particle balance equation if the diffusing particles respond to the force field at the start of the waiting time prior to jumping.
These derivations, and further extensions to include reactions, are described in greater detail, for chemotactic forcing, in a related publication
\cite{LH2010}.

Our starting point is the  generalized master equation approach developed in  \cite{Chechkin2005,SK2006}.
This approach utilizes two balance conditions.
The balance equation for  the concentration of particles, $n_i(t)$, at the site $i$ and time $t$ is
\begin{equation}
\label{Model:III:Balance}
\frac{dn_i(t)}{dt} = J_{i}^{+}(t)-J_{i}^{-}(t),
\end{equation} 
where $J_{i}^{\pm}$ are the gain (+) and loss (-) fluxes at the site $i$.
The second balance equation is a conservation equation for the arriving flux of particles at the point $i$. In general, to allow for biased CTRWs in a space-and-time-dependent force field, we write
\begin{equation}
\label{Model:III:Gain:flux}
J_{i}^{+}(t)=  p_{r}(x_{i-1},t)J_{i-1}^{-}(t)+ p_{l}(x_{i+1},t)J_{i+1}^{-}(t),
\end{equation} 
where $p_r(x,t)$ and $p_l(x,t)$ are the probabilities of jumping  from $x$ to the adjacent grid point, to the right and left directions respectively.
The two balance equations can be combined to yield
\begin{equation}
\label{Model:III:Balance:2}
\frac{dn_i(t)}{dt} = p_{r}(x_{i-1},t)J_{i-1}^{-}(t)+ p_{l}(x_{i+1},t)J_{i+1}^{-}(t)-J_{i}^{-}(t).
\end{equation} 
For CTRWs with a waiting time probability density function
$\psi(t)$ the loss flux at  site $i$  is from those particles that were  originally at $i$ at $t=0$ and wait until time $t$ to leave, and those particles that arrived at an earlier time $t^{'}$ and wait until time $t$ to leave, hence \cite{Chechkin2005}
\begin{equation}
\label{Model:III:Loss:flux}
J_{i}^{-}(t)= \psi(t)n_{i}(0)+\int\limits_{0}^{t} \psi(t-t^{'}) J_{i}^{+}(t^{'}) \; dt^{'}.
\end{equation} 
We
can combine Eq.~(\ref{Model:III:Balance}) and Eq.~(\ref{Model:III:Loss:flux}) to obtain
\begin{equation}
\label{Model:III:Loss:flux:sub}
J_{i}^{-}(t)= \psi(t)n_{i}(0)+\int\limits_{0}^{t} \psi(t-t^{'}) \left[J_{i}^{-}(t^{'})+\frac{dn_i(t^{'})}{dt}\right] \; dt^{'},
\end{equation} 
and then 
\begin{equation}
\label{Model:III:Loss:flux:lapl}
\widehat{J}_{i}^{-}(s)= \widehat{\psi}(s)n_{i}(0)+ \widehat{\psi}(s) \left[\widehat{J}_{i}^{-}(s)+s\widehat{n}_i(s)-n_{i}(0)\right],
\end{equation} 
where the hat denotes a Laplace transform with respect to time and $s$ is the Laplace transform variable. This simplifies further to 
\begin{equation}
\label{Model:III:Loss:flux:lapl:2}
\widehat{J}_{i}^{-}(s)= \frac{\widehat{\psi}(s)}{\widehat{\Phi}(s)}\widehat{n}_i(s),
\end{equation} 
where $\widehat\Phi(s)$ is the Laplace transform of the survival
probability
\begin{equation}
\Phi(t)=\int_t^\infty\psi(t')\, dt'.
\end{equation}
In the CTRW model, subdiffusion originates from a heavy-tailed waiting-time density with long-time behaviour \cite{MK2000}
\begin{equation}
\label{heavy:tail} \psi(t)\sim \frac{\kappa}{\tau}\left(\frac{t}{\tau}\right)^{-1-\alpha} 
\end{equation}
where $\alpha$  is
the anomalous exponent, $\tau$ is the characteristic waiting-time, and $\kappa$ is a dimensionless constant.
Using a Tauberian (Abelian) theorem  for small $s$ \cite{Margolin2004} 
\begin{equation}
\label{L:Ratio:ht}
 \frac{\widehat{\psi}(s)}{\widehat{\Phi}(s)} \sim A_\alpha
\frac{s^{1-\alpha}}{\tau^{\alpha}}
\end{equation}
where $A_\alpha=\frac{\alpha}{\kappa\alphaFn{1-\alpha}}$.
In the 
special case of a Mittag-Leffler waiting time density \cite{SGMR03}
the ratio, Eq.({\ref{L:Ratio:ht}), is exact and $A_\alpha=1$.
The loss flux can now be obtained by using the ratio, Eq(\ref{L:Ratio:ht}), in Eq.(\ref{Model:III:Loss:flux:lapl:2})
 and inverting the Laplace transform. Noting that the Laplace Transform of a Riemann-Liouville fractional derivative of order $\alpha$, where $0<\alpha\le 1$, is given by \cite{Podlubny1999}
\begin{equation}
\label{Laplace:FracDeriv} 
\Laplace{\RLderivOM{\alpha}{f(t)}}{s} = s^{\alpha} \widehat{f}(s)- 
\left[\RLderivOM{\alpha-1}{f(t)}\right|_{t=0}
\end{equation} 
this yields 
\begin{equation}
\label{Model:III:Loss:flux:invert}
{J}_{i}^{-}(t)= \frac{A_{\alpha}}{\tau^\alpha}\RLderivOM{1-\alpha}{n_i(t)},
\end{equation} 
where we have assumed that the last term in Eq. (\ref{Laplace:FracDeriv})
is zero.
Using this result in Eq.(\ref{Model:III:Balance:2}) yields,
\begin{eqnarray}
\label{Model:III:Balance:3}
\frac{dn_i(t)}{dt} &=&\frac{A_{\alpha}}{\tau^\alpha}\left\lbrace
                    p_{r}(x_{i-1},t)\RLderivOM{1-\alpha}{n_{i-1}(t)}\right.\nonumber\\
                    & & \left.+
                    p_{l}(x_{i+1},t)\RLderivOM{1-\alpha}{n_{i+1}(t)}-
                    \RLderivOM{1-\alpha}{n_i(t)}\right\rbrace.
\end{eqnarray} 

The jump probabilities are biased by the external space-and-time-dependent force. Here we consider (near thermodynamic equilibrium) Boltzmann weights with
\begin{equation}
p_r(x_i,t)=C\frac{\exp(-\beta V(x_{i+1},t))}{\exp(-\beta V(x_i,t))},
\end{equation}
and
\begin{equation}
p_l(x_i,t)=C\frac{\exp(-\beta V(x_{i-1},t))}{\exp(-\beta V(x_i,t))}.
\end{equation}
The jump probabilities are determined at the end of the waiting time,
when the particle must jump, so that 
\begin{equation}\label{norm}
p_r(x_i,t)+p_l(x_i,t)=1,
\end{equation}
 which defines $C$, and then
\begin{equation}
p_l(x_i,t)-p_r(x_i,t)=\frac{e^{-\beta V(x_{i-1},t)}-e^{-\beta V(x_{i+1},t)}}
{e^{-\beta V(x_{i-1},t)}+e^{-\beta V(x_{i+1},t)}}.\label{diffp}
\end{equation}

The spatial continuum limit of Eq.(\ref{Model:III:Balance:3})
can now be obtained  
by setting  $x_{i}=x$ and $x_{i\pm 1}=x\pm \Delta x$ and carrying out Taylor series expansions in $x$.
Retaining terms to order $(\Delta x)^2$ and using Eq.(\ref{norm}) yields
 \begin{eqnarray}
\frac{\partial n(x,t)}{\partial t}&=&\frac{A_\alpha}{\tau^\alpha}\left\{
\Delta x\frac{\partial}{\partial x}\left[(p_l(x,t)-p_r(x,t))\,\leftidx{_0}{D^{1-\alpha}_t} n(x,t)\right]\right.\nonumber\\
& & \left.+\frac{\Delta x^2}{2}\frac{\partial^2}{\partial x^2}\,\leftidx{_0}{D^{1-\alpha}_t} n(x,t)\right\}.\label{result1}
\end{eqnarray}
The Taylor series expansion of Eq.(\ref{diffp}) yields
\begin{equation}
p_l(x,t)-p_r(x,t)\approx \beta\Delta x \frac{\partial V(x,t)}{\partial x}+O(\Delta x^3),
\end{equation}
and then
\begin{eqnarray}
 \frac{\partial n(x,t)}{\partial t}&=&\frac{A_\alpha\beta\Delta x^2}{\tau^\alpha}\frac{\partial}{\partial x}\left[\frac{\partial V(x,t)}{\partial x}\,\leftidx{_0}{D^{1-\alpha}_t} n(x,t)\right]\nonumber\\
 & & +\frac{A_\alpha\Delta x^2}{2\tau^\alpha}\frac{\partial^2}{\partial x^2}\,\leftidx{_0}{D^{1-\alpha}_t }n(x,t)
+O(\Delta x^4).
\end{eqnarray} 
In the limit $\Delta\rightarrow 0$ and $\tau\rightarrow 0$,
with
$
\kappa_\alpha=\frac{A_\alpha\Delta x^2}{2\tau^\alpha},
$
and
$\eta_\alpha=(2\beta\kappa_\alpha)^{-1}$
we recover the fractional Fokker-Planck equation, Eq.(\ref{xtffp}),
for subdiffusion in an external space-and-time-dependent force field
\begin{equation}
F(x,t)=-\frac{\partial V(x,t)}{\partial x}.
\end{equation}

The fractional Fokker-Planck equation, Eq.(\ref{xtffp}), can also be derived from the subordinated stochastic Langevin equation motivated by physical arguments 
\cite{WMW2008} to model subdiffusion in a space-and-time-dependent force field. This representation can be formulated as a system of stochastic equations,
\begin{equation}\label{amp}
\left(\begin{array}{c}
dY_t\cr
dZ_t\cr
\end{array}\right)=\left(\begin{array}{c}
F(Y_t,Z_t)\eta^{-1}\cr
0\cr
\end{array}\right)dt+
\left(\begin{array}{c}
(2\kappa)^{1/2} dB_t\cr
dU_t\cr
\end{array}\right), 
\end{equation}
where $B_t$ is a one-dimensional Brownian motion and $U_t$ is a $\alpha$-stable L\'evy subordinator in $[0,\infty), 0<\alpha<1$.  It is asumed that
$B_t$ and $U_t$ are independent stochastic processes and the initial condition
is $Y_0=Z_0=0$. The stochastic process representing subdiffusion in a space-and-time-dependent force field is postulated to be given by \cite{WMW2008} $X_t=Y(S_t)$
where for $t\ge 0$, $S_t$ is the random time the process $U_t$ exceeds $t$.

The stochastic differential equation, Eq.(\ref{amp}), belongs to the general class of stochastic processes driven by L\'evy noise 
\cite[eq.(6.12)]{A2009}. The infinitesimal generator for the process
Eq.(\ref{amp}) is then given by \cite[eq.(6.42)]{A2009}
\begin{equation}
\begin{split}
 &A f(y,z)=\frac{F(y,z)}{\eta}\frac{\partial}{\partial y}f(y,z)
+ \kappa\frac{\del^2} {\del y^2}f(y,z)
\\ &~~~+ \int_0^\infty \left[f(y,z+z') - f(y,z)\right] \frac{\alpha}{\Gamma(1-\alpha)} z'^{-1-\alpha}dz'.
\end{split}
\end{equation}
The Fokker-Planck evolution equation
for the probability density $q_t(y,z)$ of the process $(Y_t,Z_t)$
% , Eq.(\ref{amp}),  
is given by
 % \cite[eq.(3.24)]{A2009}
\begin{align}\label{qpdf}
\frac{\del}{\del t}q_t(y,z) = A^\dagger q_t(y,z)
\end{align}
where
\begin{equation}
\begin{split}\label{adjoint}
 A^\dagger f(y,z) &= \kappa\frac{\partial^2}{\partial y^2}f(y,z)
-\frac{\partial}{\partial y}\left(\frac{F(y,z)}{\eta}f(y,z)\right)
\\&~~~-\leftidx{_0} {D_z^\alpha} f(y,z)
\end{split}
\end{equation}
defines the operator adjoint to $A$.
%\paragraph{Last exit of $(Y_t,Z_t)$ from $\mathbb R\times[0,t]$.}

Now we relate the densities $p_t(x)$ and $q_t(y,z)$ of the stochastic processes $X_t$ and $(Y_t,Z_t)$ respectively.
We write $\omega$ for a particular (random) path of the latter process, and note that the coordinates at time $t$, $(Y_t(\omega),Z_t(\omega))$ and $X_t(\omega)$, are functions of $\omega$. For a fixed interval $I$ we define the indicator function 
\begin{equation}
\delta_I(x) = \begin{cases}
               1 \text{ if } x \in I \\ 0 \text{ if } x \notin I
              \end{cases}
\end{equation}
together with the auxiliary function
\begin{widetext}
\begin{equation}
 H(t',\omega,\Delta z) =\left\{ 
\begin{array}{ll}
\delta_I(Y_{t'}(\omega))&\mbox{if}\quad Z_{t'-}(\omega)\leq t \leq Z_{t'-}(\omega)+\Delta z\\
0&\mbox{otherwise.}
                 \end{array}\right.\label{bigH}
\end{equation}
\end{widetext}
With the above notation we can write
\begin{equation}
\int_I p_t(x)\, dx=\langle \delta_I(X_t(\omega))\rangle\label{peq1}
\end{equation}
where the angle brackets represent an ensemble average over all paths $\omega$.
Given that $X_t$ can be interpreted as the $Y$ coordinate of the last position of the Markov process $(Y,Z)$ before it exits the set $\mathbb{R}\times [0,t]$, we have 
\begin{equation}
 \delta_I(X_t(\omega)) = \sum_{t' > 0}H\left(t',\omega,\Delta Z_{t'}(\omega)\right),\label{peq2}
\end{equation}
since all summands equal zero except for $t' = S_t$, in which case $Y_{t'}(\omega)=X_t(\omega)$.
The jumps $\Delta z=\Delta Z_{t'}(\omega)$ are a Poisson point process on $(0,\infty)$ whose characteristic measure has the density $\frac{\alpha}{\Gamma(1-\alpha)}\Delta z^{-1-\alpha}$.
We can now combine Eqs.(\ref{peq1}), ({\ref{peq2}) and 
use the compensation formula  in \cite[XII~(1.10)]{RY1999}, 
to write
\begin{equation}
\int_I p_t(x)dx 
= \left\langle \int_0^\infty \int_0^\infty  H(t',\omega,\Delta z) \frac{\alpha \Delta z^{-1-\alpha}}{\Gamma(1-\alpha)} d\Delta z\, dt'\right\rangle.
\end{equation}
After integrating over $\Delta z$ and using Eq.(\ref{bigH}) the right hand side simplifies further to\begin{equation}
\left\langle \int_0^\infty dt' \delta_I(Y_{t'}(\omega))\delta_{[0,t]}(Z_{t'}(\omega)) \frac{(t-Z_{t'}(\omega))^{-\alpha}}{\Gamma(1-\alpha)}\right\rangle.
\end{equation}
The ensemble average is evaluated using the probability density $q_t(y,z)$ so that
$$
 \int_I p_t(x)dx =\int_0^\infty dt' \int_I dy \int_0^t dz ~ q_{t'}(y,z) \frac{(t-z)^{-\alpha}}{\Gamma(1-\alpha)},
$$
and thus
\begin{equation}
p_t(x)=\int_0^\infty dt'~ \leftidx{_0} {I^{1-\alpha}_t} q_{t'}(x,t),
\end{equation}
where $\leftidx{_0} {I^{1-\alpha}_t}$ denotes the Liouville fractional integral of order $1-\alpha$ acting on $t$.
It also follows that
\begin{eqnarray}
\leftidx{_0} {D_t^{1-\alpha}} p_t(x)&=& \int_0^\infty q_{t'}(x,t)dt',\label{pq1}\\
\frac{\partial}{\partial t} p_t(x)&=&\int_0^\infty
\leftidx{_0} {D_t^\alpha}q_{t'}(x,t)dt'.\label{pq2}
\end{eqnarray}
We now solve Eqs.(\ref{qpdf}), (\ref{adjoint}) for $\leftidx{_0} {D_t^\alpha}q_{t'}(x,t)$
and subsitute this into Eq.(\ref{pq2}) to obtain
\begin{eqnarray}
\frac{\partial}{\partial t}p_t(x)&=& \int_0^\infty
\left(\kappa\frac{\partial^2}{\partial x^2} q_{t'}(x,t)
- \frac{1}{\eta}\frac{\partial}{\partial x}(F(x,t) q_{t'}(x,t)) \right.\nonumber
\\& & \left.~~~- \frac{\partial}{\partial {t'}} q_{t'}(x,t) \right)d{t'}
\end{eqnarray}
and finally, using Eq.(\ref{pq1}),
\begin{equation}
\frac{\partial}{\partial t}p_t(x)= \kappa\frac{\partial^2}{\partial x^2} \leftidx{_0} {D_t^{1-\alpha}} p_t(x)
- \frac{\partial}{\partial x} \left(\frac{F(x,t)}{\eta} \leftidx{_0} {D_t^{1-\alpha}} p_t(x)\right)\label{final}
\end{equation}
where we have used $q_\infty(x,t) = 0$ and $q_0(x,t) = \delta_{(0,0)}(x,t)$ and assumed $t>0$. 
Equation (\ref{final}) recovers the fractional Fokker-Planck equation for space-and-time-dependent forces, Eq.(\ref{xtffp}).

A different fractional Fokker-Planck equation can be obtained from the following
generalization of the CTRW particle balance equation
 \begin{eqnarray}
\label{ModelII:1}
n_{i}(t) &=& n_i(0) \Phi(t) + \int\limits_0^{t} \left\lbrace 
p_r(x_{i-1},t^{'}) n_{i-1}(t^{'})\right.\nonumber\\
& &\left.+
p_l(x_{i+1},t^{'}) n_{i+1}(t^{'}) \right\rbrace \psi(t-t^{'}) \:dt^{'},
\end{eqnarray} 
where the jump probabilities are evaluated
at the start of the waiting time  prior to jumping. Note that this equation can be derived from the Montroll-Weiss CTRW formalism (see e.g., \cite{HLS2010}) if and only if the jumping probabilities are independent of time. Our inclusion of time dependence
is an {\em ad-hoc} generalization for time dependent jumps.
Using Laplace transform methods  as above then leads to the discrete space evolution equation
\begin{eqnarray}
\label{ModelII:8}
\frac{d n_i}{dt} &=& \frac{A_{\alpha}}{\tau^\alpha} \RLderivO{1-\alpha} \left\lbrace  -n_i(t)+
p_r(x_{i-1},t) n_{i-1}(t)\right.\nonumber\\
& & \left.+p_l(x_{i+1},t) n_{i+1}(t)\right\rbrace.
\end{eqnarray} 
It follows from Eqs. (16) and (17) that this evolution equation is the evolution equation for the loss flux in the generalized mater equation approach.
After taking the spatial continuum limit of Eq.(\ref{ModelII:8}) with Boltzmann weighted jumping
probabilities we have
 \begin{equation}
\label{ModelII:eqn:final}
\frac{\partial n}{\partial t} = \FracderivN{\alpha} \left[ 
\kappa_{\alpha}\frac{\partial^{2}n(x,t)}{\partial x^2}+\frac{1}{\eta_\alpha}_{\alpha}\frac{\partial}{\partial x}\left(\frac{\partial V(x,t)}{\partial x} n(x,t)\right) \right].
\end{equation} 
This provides an interpretation of the
fractional Fokker-Planck equation
\begin{equation}
\frac{\partial P}{\partial t}= \leftidx{_0}{D_t^{1-\alpha}}
\left[\kappa_\alpha\frac{\partial^2}{\partial x^2}-\frac{1}{\eta_\alpha}\frac{\partial}{\partial x}
F(x,t)\right]P(x,t)
\end{equation}
as an equation for the loss flux in subdiffusion in a space-and-time-dependent force field

It is straightforward \cite{LH2010} to obtain numerical solutions of the discrete space evolution equations for subdiffusion in  space-and-time-dependent force fields, Eqs.(\ref{Model:III:Balance:3}), (\ref{ModelII:8}) using  an implicit time stepping method with the fractional derivatives approximated using the L1 scheme \cite{Oldham1974}. The CTRW models described here are also easy to simulate using
Monte Carlo methods \cite{LH2010}.
%As an example we consider subdiffusion in the space-and-time-dependent field with potential
%\begin{equation}
%V(x,t)=\frac{1}{4}x^4-\frac{1}{2}x^2-\epsilon x\cos t.\label{VXT}
%\end{equation}
%The standard Fokker-Planck equation for this and related potentials has been widely studied as a model for stochastic resonance with additive Gaussian noise \cite{GHJM1998}. 
\begin{acknowledgements}
This work was supported by the Australian Research Council. 
We are grateful to Eli Barkai
for his comments on our draft manuscript.
\end{acknowledgements}

\bibliographystyle{apsrev}
\bibliography{HLS-PRL} 

\begin{thebibliography}{22}
\expandafter\ifx\csname natexlab\endcsname\relax\def\natexlab#1{#1}\fi
\expandafter\ifx\csname bibnamefont\endcsname\relax
  \def\bibnamefont#1{#1}\fi
\expandafter\ifx\csname bibfnamefont\endcsname\relax
  \def\bibfnamefont#1{#1}\fi
\expandafter\ifx\csname citenamefont\endcsname\relax
  \def\citenamefont#1{#1}\fi
\expandafter\ifx\csname url\endcsname\relax
  \def\url#1{\texttt{#1}}\fi
\expandafter\ifx\csname urlprefix\endcsname\relax\def\urlprefix{URL }\fi
\providecommand{\bibinfo}[2]{#2}
\providecommand{\eprint}[2][]{\url{#2}}

\bibitem[{\citenamefont{Metzler and Klafter}(2000)}]{MK2000}
\bibinfo{author}{\bibfnamefont{R.}~\bibnamefont{Metzler}} \bibnamefont{and}
  \bibinfo{author}{\bibfnamefont{J.}~\bibnamefont{Klafter}},
  \bibinfo{journal}{Phys. Rep.} \textbf{\bibinfo{volume}{339}},
  \bibinfo{pages}{1} (\bibinfo{year}{2000}).

\bibitem[{\citenamefont{B.{ I. Henry} et~al.}(2010)\citenamefont{B.{ I. Henry},
  T.{ A.M. Langlands}, and Straka}}]{HLS2010}
\bibinfo{author}{\bibnamefont{B.{ I. Henry}}},
  \bibinfo{author}{\bibnamefont{T.{ A.M. Langlands}}}, \bibnamefont{and}
  \bibinfo{author}{\bibfnamefont{P.}~\bibnamefont{Straka}}, in
  \emph{\bibinfo{booktitle}{Complex Physical, Biophysical and Econophysical
  Systems: World Scientific Lecture Notes in Complex Systems}}, edited by
  \bibinfo{editor}{\bibfnamefont{R.}~\bibnamefont{{L. Dewar}}}
  \bibnamefont{and} \bibinfo{editor}{\bibfnamefont{F.}~\bibnamefont{Detering}}
  (\bibinfo{publisher}{World Scientific}, \bibinfo{address}{Singapore},
  \bibinfo{year}{2010}), vol.~\bibinfo{volume}{9}, pp. \bibinfo{pages}{37--90}.

\bibitem[{\citenamefont{Metzler
  et~al.}(1999{\natexlab{a}})\citenamefont{Metzler, Barkai, and
  Klafter}}]{MBK1999b}
\bibinfo{author}{\bibfnamefont{R.}~\bibnamefont{Metzler}},
  \bibinfo{author}{\bibfnamefont{E.}~\bibnamefont{Barkai}}, \bibnamefont{and}
  \bibinfo{author}{\bibfnamefont{J.}~\bibnamefont{Klafter}},
  \bibinfo{journal}{Europhys Letts} \textbf{\bibinfo{volume}{46}},
  \bibinfo{pages}{431} (\bibinfo{year}{1999}{\natexlab{a}}).

\bibitem[{\citenamefont{Barkai et~al.}(2000)\citenamefont{Barkai, Metzler, and
  Klafter}}]{BMK2000}
\bibinfo{author}{\bibfnamefont{E.}~\bibnamefont{Barkai}},
  \bibinfo{author}{\bibfnamefont{R.}~\bibnamefont{Metzler}}, \bibnamefont{and}
  \bibinfo{author}{\bibfnamefont{J.}~\bibnamefont{Klafter}},
  \bibinfo{journal}{Phys. Rev. E} \textbf{\bibinfo{volume}{61}},
  \bibinfo{pages}{132} (\bibinfo{year}{2000}).

\bibitem[{\citenamefont{Metzler
  et~al.}(1999{\natexlab{b}})\citenamefont{Metzler, Barkai, and
  Klafter}}]{MBK1999a}
\bibinfo{author}{\bibfnamefont{R.}~\bibnamefont{Metzler}},
  \bibinfo{author}{\bibfnamefont{E.}~\bibnamefont{Barkai}}, \bibnamefont{and}
  \bibinfo{author}{\bibfnamefont{J.}~\bibnamefont{Klafter}},
  \bibinfo{journal}{Phys. Rev. Letts.} \textbf{\bibinfo{volume}{82}},
  \bibinfo{pages}{3563} (\bibinfo{year}{1999}{\natexlab{b}}).

\bibitem[{\citenamefont{Montroll and Weiss}(1965)}]{MW1965}
\bibinfo{author}{\bibfnamefont{E.}~\bibnamefont{Montroll}} \bibnamefont{and}
  \bibinfo{author}{\bibfnamefont{G.}~\bibnamefont{Weiss}}, \bibinfo{journal}{J.
  Math. Phys.} \textbf{\bibinfo{volume}{6}}, \bibinfo{pages}{167}
  (\bibinfo{year}{1965}).

\bibitem[{\citenamefont{I.{ M. Sokolov} and Klafter}(2006)}]{SK2006}
\bibinfo{author}{\bibnamefont{I.{ M. Sokolov}}} \bibnamefont{and}
  \bibinfo{author}{\bibfnamefont{J.}~\bibnamefont{Klafter}},
  \bibinfo{journal}{Phys. Rev. Lett.} \textbf{\bibinfo{volume}{97}},
  \bibinfo{pages}{140602} (\bibinfo{year}{2006}).

\bibitem[{\citenamefont{Heinsalu et~al.}(2007)\citenamefont{Heinsalu,
  Patriarca, Goychuk, and H\"{a}nggi}}]{HPGH2007}
\bibinfo{author}{\bibfnamefont{E.}~\bibnamefont{Heinsalu}},
  \bibinfo{author}{\bibfnamefont{M.}~\bibnamefont{Patriarca}},
  \bibinfo{author}{\bibfnamefont{I.}~\bibnamefont{Goychuk}}, \bibnamefont{and}
  \bibinfo{author}{\bibfnamefont{P.}~\bibnamefont{H\"{a}nggi}},
  \bibinfo{journal}{Phys. Rev. Lett.} \textbf{\bibinfo{volume}{99}},
  \bibinfo{pages}{120602} (\bibinfo{year}{2007}).

\bibitem[{\citenamefont{Heinsalu et~al.}(2009)\citenamefont{Heinsalu,
  Patriarca, Goychuk, and P.{ H\"anggi}}}]{HPGH2009}
\bibinfo{author}{\bibfnamefont{E.}~\bibnamefont{Heinsalu}},
  \bibinfo{author}{\bibfnamefont{M.}~\bibnamefont{Patriarca}},
  \bibinfo{author}{\bibfnamefont{I.}~\bibnamefont{Goychuk}}, \bibnamefont{and}
  \bibinfo{author}{\bibnamefont{P.{ H\"anggi}}}, \bibinfo{journal}{Phys. Rev.
  E} \textbf{\bibinfo{volume}{79}}, \bibinfo{pages}{041137}
  (\bibinfo{year}{2009}).

\bibitem[{\citenamefont{Weron et~al.}(2008)\citenamefont{Weron, Magdziarz, and
  Weron}}]{WMW2008}
\bibinfo{author}{\bibfnamefont{A.}~\bibnamefont{Weron}},
  \bibinfo{author}{\bibfnamefont{M.}~\bibnamefont{Magdziarz}},
  \bibnamefont{and} \bibinfo{author}{\bibfnamefont{K.}~\bibnamefont{Weron}},
  \bibinfo{journal}{Phys. Rev. E} \textbf{\bibinfo{volume}{77}},
  \bibinfo{pages}{036704} (\bibinfo{year}{2008}).

\bibitem[{\citenamefont{Y.{-M. Kang} and Y.{-L . Jiang}}(2010)}]{KJ2010}
\bibinfo{author}{\bibnamefont{Y.{-M. Kang}}} \bibnamefont{and}
  \bibinfo{author}{\bibnamefont{Y.{-L . Jiang}}}, \bibinfo{journal}{J. Math.
  Phys.} \textbf{\bibinfo{volume}{51}}, \bibinfo{pages}{{023301}}
  (\bibinfo{year}{2010}).

\bibitem[{\citenamefont{Y.{-M. Kang} and Y.{-L . Jiang}}(2008)}]{MWK2008}
\bibinfo{author}{\bibnamefont{Y.{-M. Kang}}} \bibnamefont{and}
  \bibinfo{author}{\bibnamefont{Y.{-L . Jiang}}}, \bibinfo{journal}{Phys. Rev.
  Letts.} \textbf{\bibinfo{volume}{101}}, \bibinfo{pages}{{210601}}
  (\bibinfo{year}{2008}).

\bibitem[{\citenamefont{Kalmykov et~al.}(2006)\citenamefont{Kalmykov, Coffey,
  and Titov}}]{KYC2006}
\bibinfo{author}{\bibfnamefont{Y.~P.} \bibnamefont{Kalmykov}},
  \bibinfo{author}{\bibfnamefont{W.~T.} \bibnamefont{Coffey}},
  \bibnamefont{and} \bibinfo{author}{\bibfnamefont{S.~V.} \bibnamefont{Titov}},
  \bibinfo{journal}{Phys. Rev. E} \textbf{\bibinfo{volume}{74}},
  \bibinfo{pages}{011105} (\bibinfo{year}{2006}).

\bibitem[{\citenamefont{{M. Mousa} and Kaltayev}(2009)}]{MK2009}
\bibinfo{author}{\bibfnamefont{M.}~\bibnamefont{{M. Mousa}}} \bibnamefont{and}
  \bibinfo{author}{\bibfnamefont{A.}~\bibnamefont{Kaltayev}},
  \bibinfo{journal}{Z. Naturforsch.} \textbf{\bibinfo{volume}{64a}},
  \bibinfo{pages}{788} (\bibinfo{year}{2009}).

\bibitem[{\citenamefont{{A.M. Langlands} and {I. Henry}}(2010)}]{LH2010}
\bibinfo{author}{\bibfnamefont{T.}~\bibnamefont{{A.M. Langlands}}}
  \bibnamefont{and} \bibinfo{author}{\bibfnamefont{B.}~\bibnamefont{{I.
  Henry}}}, \bibinfo{journal}{arXiv:submit/0000822} pp. \bibinfo{pages}{1--25}
  (\bibinfo{year}{2010}).

\bibitem[{\citenamefont{A.{V. Chechkin} et~al.}(2005)\citenamefont{A.{V.
  Chechkin}, Gorenflo, and I.{M. Sokolov}}}]{Chechkin2005}
\bibinfo{author}{\bibnamefont{A.{V. Chechkin}}},
  \bibinfo{author}{\bibfnamefont{R.}~\bibnamefont{Gorenflo}}, \bibnamefont{and}
  \bibinfo{author}{\bibnamefont{I.{M. Sokolov}}}, \bibinfo{journal}{J. Phys. A:
  Math. Gen.} \textbf{\bibinfo{volume}{38}}, \bibinfo{pages}{L679}
  (\bibinfo{year}{2005}).

\bibitem[{\citenamefont{Margolin}(2004)}]{Margolin2004}
\bibinfo{author}{\bibfnamefont{G.}~\bibnamefont{Margolin}},
  \bibinfo{journal}{Physica A} \textbf{\bibinfo{volume}{334}},
  \bibinfo{pages}{46} (\bibinfo{year}{2004}).

\bibitem[{\citenamefont{Scalas et~al.}(2003)\citenamefont{Scalas, Gorenflo,
  Mainardi, and Raberto}}]{SGMR03}
\bibinfo{author}{\bibfnamefont{E.}~\bibnamefont{Scalas}},
  \bibinfo{author}{\bibfnamefont{R.}~\bibnamefont{Gorenflo}},
  \bibinfo{author}{\bibfnamefont{F.}~\bibnamefont{Mainardi}}, \bibnamefont{and}
  \bibinfo{author}{\bibfnamefont{M.}~\bibnamefont{Raberto}},
  \bibinfo{journal}{Fractals} \textbf{\bibinfo{volume}{11}},
  \bibinfo{pages}{281} (\bibinfo{year}{2003}).

\bibitem[{\citenamefont{Podlubny}(1999)}]{Podlubny1999}
\bibinfo{author}{\bibfnamefont{I.}~\bibnamefont{Podlubny}},
  \emph{\bibinfo{title}{Fractional differential equations}}, vol.
  \bibinfo{volume}{198} of \emph{\bibinfo{series}{Mathematics in Science and
  Engineering}} (\bibinfo{publisher}{Academic Press}, \bibinfo{address}{New
  York and London}, \bibinfo{year}{1999}).

\bibitem[{\citenamefont{Applebaum}(2009)}]{A2009}
\bibinfo{author}{\bibfnamefont{D.}~\bibnamefont{Applebaum}},
  \emph{\bibinfo{title}{Levy Processes and Stochastic Calculus, 2nd Edition}},
  vol. \bibinfo{volume}{116} of \emph{\bibinfo{series}{Cambridge Studies in
  Advanced Mathematics}} (\bibinfo{publisher}{Cambridge University Press},
  \bibinfo{address}{Cambridge}, \bibinfo{year}{2009}).

\bibitem[{\citenamefont{Revuz and Yor}(1999)}]{RY1999}
\bibinfo{author}{\bibfnamefont{D.}~\bibnamefont{Revuz}} \bibnamefont{and}
  \bibinfo{author}{\bibfnamefont{M.}~\bibnamefont{Yor}},
  \emph{\bibinfo{title}{Continuous Martingales and Brownian Motion, 3rd
  Edition}}, vol. \bibinfo{volume}{293} of \emph{\bibinfo{series}{A Series of
  Comprehensive Studies in Mathematics}} (\bibinfo{publisher}{Springer},
  \bibinfo{address}{Berlin}, \bibinfo{year}{1999}).

\bibitem[{\citenamefont{Oldham and Spanier}(1974)}]{Oldham1974}
\bibinfo{author}{\bibfnamefont{K.}~\bibnamefont{Oldham}} \bibnamefont{and}
  \bibinfo{author}{\bibfnamefont{J.}~\bibnamefont{Spanier}},
  \emph{\bibinfo{title}{The Fractional Calculus: Theory and Applications of
  Differentiation and Integration to Arbitrary Order}}, vol.
  \bibinfo{volume}{111} of \emph{\bibinfo{series}{Mathematics in Science and
  Engineering}} (\bibinfo{publisher}{Academic Press}, \bibinfo{address}{New
  York and London}, \bibinfo{year}{1974}).

\end{thebibliography}

\end{document}